# Second-order nonlinear optical metamaterials: ABC-type nanolaminates

**Authors:** Luca Alloatti[1†*], Clemens Kieninger[1], Andreas Froelich[2-4], Matthias Lauermann[1], Tobias Frenzel[2], Kira Köhnle[1], Wolfgang Freude[1,5], Juerg Leuthold[6], Martin Wegener[2-4], and Christian Koos[1,3,5,**]

**Affiliations:**

[1]Institute of Photonics and Quantum Electronics (IPQ),
Karlsruhe Institute of Technology (KIT), 76128 Karlsruhe, Germany.

[2]Institute of Applied Physics, Karlsruhe Institute of Technology (KIT),
76128 Karlsruhe, Germany.

[3]DFG-Center for Functional Nanostructures (CFN), Karlsruhe Institute of Technology (KIT),

76128 Karlsruhe, Germany.

[4]Institute of Nanotechnology, Karlsruhe Institute of Technology (KIT),
76021 Karlsruhe, Germany.

[5]Institute for Microstructure Technology (IMT), Karlsruhe Institute of Technology (KIT),
76344 Eggenstein-Leopoldshafen.

[6]Institute of Electromagnetic Fields, Eidgenössische Technische Hochschule (ETH),
ETZ K 81, Gloriastrasse 35, 8092 Zürich, Switzerland.

[†] Present address: Research Laboratory of Electronics, Massachusetts Institute of Technology (MIT), 77 Massachusetts Ave. 36-477, Cambridge MA 02139, USA

[*]alloatti@mit.edu; [**]christian.koos@kit.edu

**Structuring optical materials on a nanometer scale can lead to artificial effective media, or metamaterials, with strongly altered optical behavior. Metamaterials can provide a wide range of linear optical properties such as negative refractive index[1,2], hyperbolic dispersion[3], or magnetic behavior at optical frequencies[4]. Nonlinear optical properties, however, have only been demonstrated for patterned metallic films[5-10] which suffer from high optical losses[11]. Here we show that second-order nonlinear metamaterials can also be obtained from non-metallic centrosymmetric constituents with inherently low optical absorption.**

**In our proof-of-principle experiments, we have iterated atomic-layer deposition (ALD) of three different constituents, A = $Al_2O_3$, B = $TiO_2$ and C = $HfO_2$. The centrosymmetry of the resulting ABC stack is broken since the ABC and the inverted CBA sequences are not equivalent - a necessary condition for non-zero second-order nonlinearity. To the best of our knowledge, this is the first realization of a bulk nonlinear optical metamaterial.**

The basic idea of metamaterials is simple, yet powerful: Using ordinary constituents shaped on a sufficiently small spatial scale, effective material properties that go qualitatively beyond those of the ingredients become possible[12,13]. Early examples are stacks of isotropic layers leading to nanolaminates with an effective anisotropic electromagnetic response[14]. Metamaterials may also lead to properties that are typically not present in nature, such as hyperbolic dispersion observed in layered metal-dielectric structures[3] or negative refractive indices which can be exploited to create superlenses[2].

However, while these examples refer to linear optical properties only, nonlinear phenomena are of fundamental technological importance as well. For example, optical nonlinearities are utilized for generating frequencies otherwise hardly accessible[7], for modulating light at hundreds of GHz[15], or for creating optical gates[16]. Nonlinear optics additionally enables fundamental components for quantum applications, such as sources of entangled photons by spontaneous down-conversion[17].

Second-order nonlinear optical crystals, such as potassium dihydrogen phosphate (KDP), lithium niobate ($LiNbO_3$) or superlattices grown by molecular-beam epitaxy (MBE)[18] are readily available, but these materials are difficult to incorporate in most photonic platforms. Nonlinear optical metamaterials are therefore interesting not only for answering general physical questions, but also because they may offer a method of incorporating second-order nonlinear materials on photonic platforms.

Current research on nonlinear optical metamaterials has shown interesting results already such as high-harmonic generation[9] or giant surface nonlinearities[5]. However, most results available so far report on enhancements of pre-existing nonlinearities which originate at metal-dielectric interfaces[5,8-10,19]. Furthermore, nonlinear metamaterials available so far consist of metallic films nanostructured into various shapes, such as split-rings[7], and are therefore associated with high optical losses[11].

In this work we present the first nonlinear metamaterials made of non-metallic constituents, and the first metamaterials with an effective bulk second-order optical nonlinearity. Our materials consist of thin films made by atomic-layer deposition (ALD) and can therefore be deposited on virtually any surface[20]. In our experiments we have achieved a nonlinear susceptibility of 0.26 pm/V (Supplementary Information).

In our demonstration, we arrange layers into a non-centrosymmetric ABC-stack such that the individual surface nonlinearities[21,22] originating at the boundary of neighboring materials do not sum up to zero. A TEM image of a typical ABC-stack is shown Fig. 1a. As constituents, we chose centrosymmetric dielectrics, A = $Al_2O_3$, B = $TiO_2$, and C = $HfO_2$, for being able to exclude any second-order nonlinear contribution from the bulk of the individual layers. These dielectrics are furthermore used off-resonance and are transparent throughout the visible and the near-infrared spectral range. We grew samples of the form $M\times(N\times A, N\times B, N\times C)$, meaning that $N$ ALD cycles of the material A were deposited first, followed by $N$ cycles of the material B, and by $N$ cycles of the material C. The procedure was then repeated $M$ times such as to obtain a total number of $3\times M\times N = 900$ cycles which was kept constant to for all samples. The numbers $N$ and $M$ were varied, so as to obtain samples with different interface densities.

Samples were grown on a glass substrate and the nonlinearity was investigated by second-harmonic generation (SHG) under fs-excitation at a pump wavelength of 800 nm (Methods). The samples were tilted by $\vartheta = 45°$ towards the incident beam, and the p-polarized pump laser was focused to a spot radius of $w = 24.5\ \mu m$, Fig. 1b. The SHG signal was collected by a lens, separated from the pump laser using dielectric filters, and finally detected by a photomultiplier (Methods). All samples exhibit a similar total thickness of about 68 nm, as confirmed independently by optical ellipsometry (Methods).

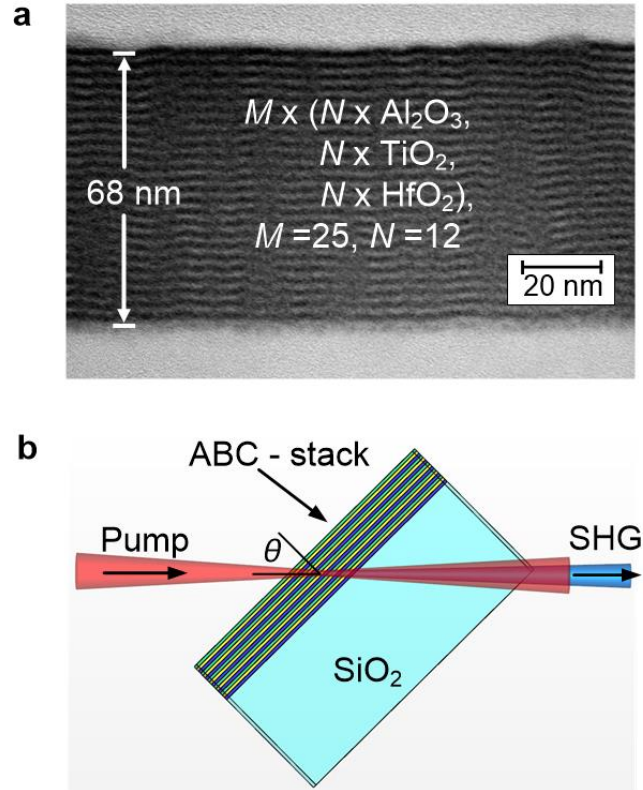

**Fig. 1 | ABC-sample cross-section and SHG representation. a,** TEM image of an ABC-stack made by iterating $M = 25$ times the deposition of $N = 12$ cycles of materials $A = Al_2O_3$, $B = TiO_2$, and $C = HfO_2$. The layered structure is clearly visible, however, since the TEM contrast between $Al_2O_3$ and $TiO_2$ is low, and since $TiO_2$ layers are five times thinner than those of the other two materials, it is difficult to distinguish $Al_2O_3$ and $TiO_2$ in the TEM image. The total number of ALD cycles is $3 \times M \times N = 900$ and correspond to a thickness of 68 nm. **b,** representation of the characterization method: the nanolaminate is grown on a glass slide and a fs-pump laser at a wavelength of 800 nm is focused at an angle $\theta$ on the nanolaminate. Second-harmonic generation (SHG) is used to measure the effective bulk second order nonlinearity of the nanolaminates.

The SHG signal for constant pump power is plotted versus $N$ in Fig. 2. The second-harmonic signal shows a maximum for $N = 12$, and decreases approximately like $1/N^2$ for increasing $N$ until the noise floor is reached. This behavior is consistent with the interpretation that the nonlinearity is proportional to the density $1/N$ of interfaces: the amplitude scales with $1/N$, the intensity with its square, and phase-matching is assumed because of the small stack thickness. The observed decrease in signal intensity for $N<12$ suggests instead that, for at least one species, ALD may not have formed a closed film (at the locations where the film is not closed, the centrosymmetry is locally recovered leading to a diminished nonlinearity), and in fact material C exhibited five times smaller growth rates per cycles (Methods and Supplementary Information).

As control experiment we grew nanolaminates made with all the three binary permutations which can be obtained using A, B and C, namely AB, BC, and AC. For better comparison with the ABC ternary samples, we chose $N = 12$ (which maximized the SHG signal for ABC samples), and $M = 38$ for a total of 912 cycles, i.e. not less than in the ternary case. Binary samples showed negligible SHG generation, consistently with the fact that they are symmetric under reflections orthogonal to the film surface. Similar control samples of pure A, B and C with at least 900 cycles for each were grown and showed negligible SHG signal indicating that nonlinear ABC stacks can be obtained by using individual linear materials. This result is implicit in Fig. 2 as well, since it corresponds to the large-$N$ limit.

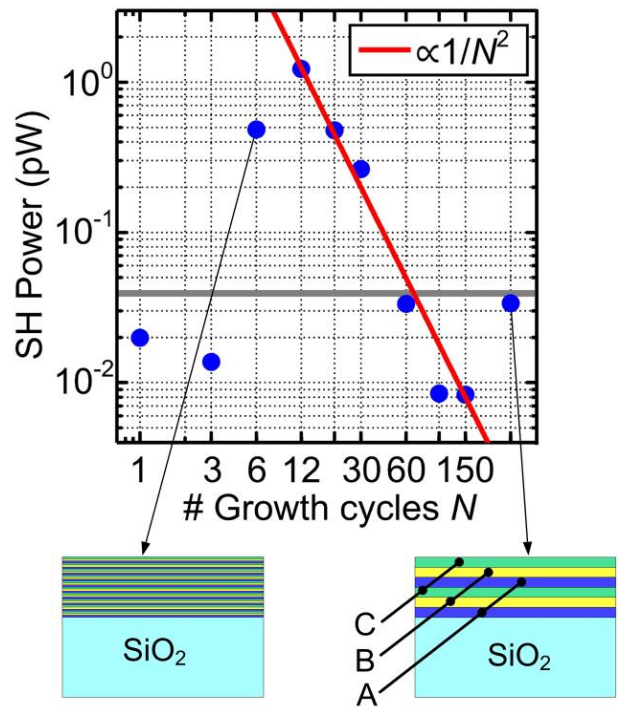

**Fig. 2 | Second-harmonic signal vs. number of consecutive deposition cycles $N$ for constant sample thickness.** The deposition of $N$ ALD cycles of A, B, and C is repeated $M$ times. The total number $3 \times M \times N$ of ALD cycles is kept constant to 900, therefore the film thickness of every sample is approximately constant, as confirmed by SEM inspection and ellipsometry (Methods). The gray line represents the SH signal from a bare glass substrate. The second-harmonic signal shows a maximum for $N = 12$, and decreases approximately like $1/N^2$ for increasing $N$ until the noise floor is reached. This behavior is consistent with the interpretation that the nonlinearity is proportional to the density $1/N$ of interfaces. The observed decrease in signal intensity for $N<12$ suggests instead that the ALD films of at least one material may not have formed a closed film. Each sample was grown in 11 hours, during five days, with a growth order meant for minimizing possible errors due to accidental variations of the growing conditions (Methods). Bottom: illustration of samples with small and large $N$.

We further verified that the SHG signal scales as expected with the square of pump power, Fig. 3a. Moreover, we grew samples of different thicknesses by varying the number of ABC periods $M = 10, 25, 50$ for constant $N = 12$. The observed SHG power scales $\propto M^2$, consistently with the interpretation of assigning a bulk-effective nonlinearity to the nanolaminates, Fig. 3b. For describing the nanolaminates as an effective bulk nonlinear medium it is further necessary that the ABC-lattice constant is much smaller than the optical wavelengths involved; for the best sample depicted in Fig. 2 ($N = 12$, $M = 25$) the ABC-lattice constant is $a = 68$ nm/$M = 2.7$ nm so that this condition is clearly fulfilled when compared to 800 nm fundamental or 400 nm second-harmonic wavelength.

For the samples exhibiting the highest nonlinearity, $N = 12$, we recorded the s- and p-polarized second-harmonic signal as a function of the pump polarization $\varphi$ (s-polarization corresponds to $\varphi = 0°$, p-polarization to $\varphi = 90°$), Fig. 3c,d. By fitting these data with the theoretical expectation for a bulk effective medium with symmetry $C_{\infty,v}$ it is possible to extract the three independent second-order tensor elements (Supplementary Information)[23], resulting to $\chi^{(2)}_{zzz} = 0.26\,\mathrm{pm/V}$, $\chi^{(2)}_{yyz} = 0.18\,\mathrm{pm/V}$, and $\chi^{(2)}_{zxx} = 0.10\,\mathrm{pm/V}$.

The nonlinearity of the nanolaminates can likely be increased in several ways. For example, it can be shown that quaternary samples can achieve higher nonlinearities than any ternary sample utilizing the same ingredients (under the assumption that the nonlinearity originates at the interfaces only). Furthermore, there is a large variety of other materials which can be grown by ALD, and it is likely that different material combinations will lead to stronger effective bulk nonlinearities. Moreover, optimized fabrication recipes will enable thinner individual layers and higher density of interfaces.

Since the nanolaminates proposed in this work can be grown on virtually any surface, they may become excellent candidates for integrating second-order nonlinearity into generic photonic platforms. Moreover, the ALD technique is widely available, is standard in CMOS foundries[24], and does not require elevated temperatures. This combination makes it compatible with a variety of existing processes, in particular with front-end-of-line fabrication of silicon circuits. Furthermore, because of the transparency of the individual ingredients in the visible spectral range, our nanolaminates may further find applications which are out of reach for the majority of photonic platforms, including silicon photonics or III-V technologies.

In conclusion, we have shown that by structuring three centrosymmetric dielectrics on a nanometer scale, the resulting nanolaminates may exhibit second-order nonlinearity. This is the first demonstration of a bulk second-order nonlinear metamaterial. The films are grown by ALD, which allows for deposition on virtually any surface. We believe that this technology may have disruptive consequences in fields such as infrared and visible nanophotonics.

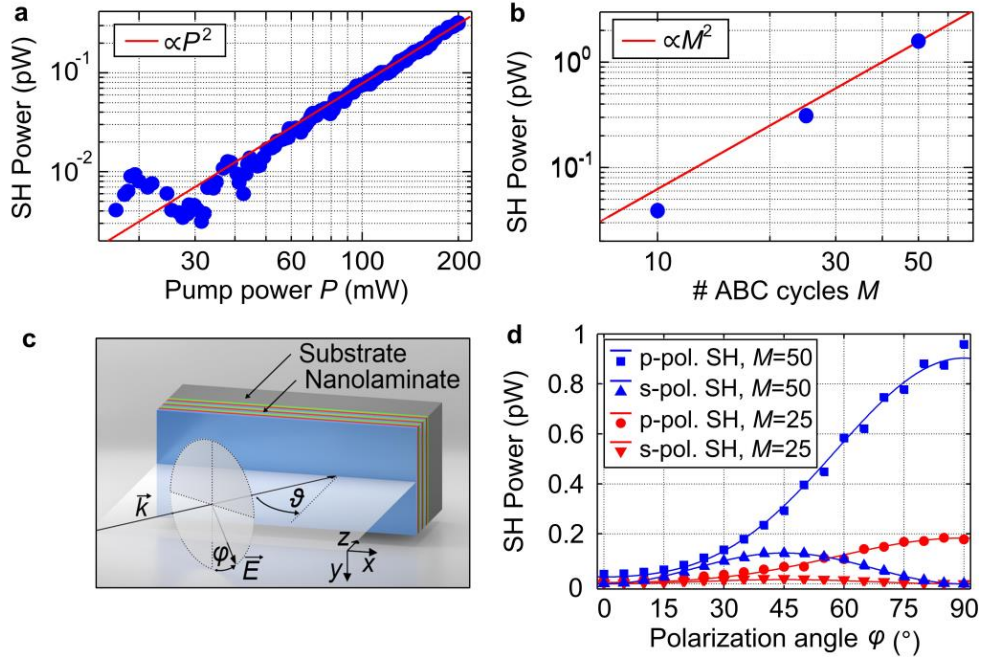

**Fig. 3 | Second-harmonic signal vs. pump power, sample thickness and polarization. a**, Measured SH power (dots) versus incident pump power $P$ for sample with $N = 12$ and $M = 25$. Parameters are: incident p-polarization, $\vartheta = 45°$. The straight line corresponds to quadratic dependence. **b,** Measured SH power (dots) vs. number of periods $M = 12, 25, 50$. All samples have $N = 12$. The straight line is the expectation for phase-matched bulk SH generation. Parameters are as in **b** and $P = 200$ mW. **c**, Illustration of the excitation geometry including the angle of incidence $\vartheta$ and the polarization angle $\varphi$. Incident s-polarization corresponds to $\varphi = 0°$, p-polarization to $\varphi = 90°$. **d**, s- and p-polarized SH power versus pump polarization angle $\varphi$ for samples with $M = 25$ and $M = 50$ (see legend). Both samples have $N = 12$, angle of incidence $\vartheta = 55°$ and $P = 160$ mW. The solid curves are calculated for a single set of $\chi^{(2)}_{ijk}$ tensor elements of an effective bulk material with the symmetry of the sample $C_{\infty v}$.

## Methods

**Sample fabrication.** All films were fabricated by atomic-layer deposition (ALD) using a Savannah 100 reactor by Cambridge Nanotech. Standard silica glass substrates with a thickness of either 175 µm or 1 mm were used. The bottom sides of the glass substrates were covered with Kapton tape to avoid deposition on this surface. The tape was removed after growth. The sample and reactor were heated to a temperature of 150°C. The precursors for Al, Ti and Hf and O were trimethylaluminum 97% (Sigma Aldrich P.Nr. 257222), Titanium(IV) isopropoxide 99.999% (Sigma-Aldrich P.Nr. 377996), Tetrakis(dimethylamido)hafnium(IV) ≥99.99% (Sigma-Aldrich P.Nr. 455199) and hydrogen peroxide 30% (Merck P.Nr. 107209), respectively. The chamber was constantly flushed with 20 sccm of Ar (unless differently specified). Further parameters are given in Table 1. For minimizing possible systematic errors, e.g., due to precursor degradation, the samples shown in Fig. 2 were grown in the following order (first to last): $N = 3, 12, 30, 100, 300, 150, 60, 20, 6, 1$. Thereafter, we grew the bulk samples ($N = 900$ growth cycles of pure A, B or C) and the binary reference samples (AB, BC and AC types).

The TEM image in Fig. 1a was taken with a CM200 system from Philips. All sample thicknesses were independently measured by optical ellipsometry. In the analysis of these data, we have assumed a single homogeneous film. For example, for the sample shown in Fig. 2, ellipsometry delivers a refractive index of 1.878 at 800 nm wavelength and a total thickness of 72 nm. Within the uncertainties, this value is consistent with the 68 nm determined by TEM. The thicknesses of the other samples as measured by optical ellipsometry are summarized in Table 2.

| Precursor for | **Aluminum** | **Titanium** | **Hafnium** | **Oxygen** |
|---|---|---|---|---|
| Temperature (°C) | not heated | 80 | 75 | not heated |
| Pulse duration (s) | 0.015 | 0.1 | 0.15 | 0.015 |
| Wait time after pulse (s) | 20 | see caption | 20 | 20 |

**Table 1. Parameters used for ALD.** Titanium was inserted with closed vacuum valve. The valve was closed before the insertion of the titanium precursor and reopened two seconds later; subsequently the chamber was flushed for twenty seconds with 100 sccm of argon.

| ABC laminate samples with $M\times(3N) = 900$ ALD growth cycles | | | | | | | | | | |
|---|---|---|---|---|---|---|---|---|---|---|
| $N$ | 1 | 3 | 6 | 12 | 20 | 30 | 60 | 100 | 150 | 300 |
| Thickness (nm) | 77 | 76 | 74 | 72 | 73 | 77 | 74 | 81 | 81 | 93 |

| Binary laminate samples with $38\times(2N) = 912$ (for $N = 12$) ALD growth cycles | | | |
|---|---|---|---|
| Constituents | A/B = $Al_2O_3$/ $TiO_2$ | B/C = $TiO_2$/ $HfO_2$ | A/C = $Al_2O_3$/ $HfO_2$ |
| Thickness (nm) | 57 | 63 | 100 |

| Bulk samples with $N = 900$ ALD growth cycles | | | |
|---|---|---|---|
| Constituent | A = $Al_2O_3$ | B = $TiO_2$ | C = $HfO_2$ |
| Thickness (nm) | 102 | 20 | 105 |

**Table 2 | Thicknesses of fabricated samples as determined by optical ellipsometry.** The tables indicate the thicknesses of ternary, binary and bulk nanolaminates (from top to bottom). For the ternary samples, the thickness remains almost unchanged for each value of $N$. For the bulk samples, $TiO_2$ exhibited 5 times smaller growth rates than $Al_2O_3$ and $HfO_2$. This is consistent with the thicknesses of the binary samples.

**Measurement setup.** To generate the second-harmonic, we excited all samples by focusing laser pulses with 165 fs duration, 80.6 MHz repetition rate, and a center wavelength of 800 nm (Tsunami, Spectra-Physics) onto the samples. Focusing was accomplished by a lens with a focal length of 20 cm.

The light emerging from the samples was filtered with one band-pass filter (central wavelength 400 nm, FWHM = 40 nm, Thorlabs FB400-40, optical density > 4 @ 800 nm, 48% transmission

@ 400 nm) and two short-pass filters (cut-on wavelength 700 nm, Thorlabs FESH0700, optical density > 5 @ 800 nm, 97% transmission @ 400 nm). When one short-pass filter was removed from the setup, no significant change in the response was detected, showing that the signal detected was not an artifact due to unfiltered pump-light. For detection of the SH signal, we use a photomultiplier Hamamatsu R4332 at a cathode voltage of -1240 V. The photomultiplier was terminated with a 50 Ω resistor. For increased sensitivity, the incident light was chopped at 177 Hz to allow for lock-in detection. For each measurement, after optimizing the focus position by maximizing the detected signal, the SH power was recorded as a function of the pump-power (varied from zero to the maximum by rotating a $\lambda/2$ plate placed before a polarization cube). As to be expected, all second-order nonlinear signals discussed in this work scaled nearly quadratically versus the fundamental input power over a significant range. For incident average powers exceeding approximately 320 mW, however, a roll-off was observed, which may be attributed to avalanche breakdown, surface contamination or sample heating via residual absorption[25].

## References

1   Shalaev, V. M. Optical negative-index metamaterials. *Nature Photon.* **1**, 41-48 (2007).

2   Pendry, J. B. Negative refraction makes a perfect lens. *Phys. Rev. Lett.* **85**, 3966-3969 (2000).

3   Poddubny, A., Iorsh, I., Belov, P. & Kivshar, Y. Hyperbolic metamaterials. *Nature Photon.* **7**, 948-957 (2013).

4   Soukoulis, C. M., Linden, S. & Wegener, M. Negative refractive index at optical wavelengths. *Science* **315**, 47-49 (2007).

5   Lee, J. *et al.* Giant nonlinear response from plasmonic metasurfaces coupled to intersubband transitions. *Nature* **511**, 65-U389 (2014).

6   Klein, M. W., Enkrich, C., Wegener, M. & Linden, S. Second-harmonic generation from magnetic metamaterials. *Science* **313**, 502-504 (2006).

7   Luo, L. *et al.* Broadband terahertz generation from metamaterials. *Nature Commun.* **5**, 3055-3055 (2014).

8   Kauranen, M. & Zayats, A. V. Nonlinear plasmonics. *Nature Photon.* **6**, 737-748 (2012).

9   Kim, S. *et al.* High-harmonic generation by resonant plasmon field enhancement. *Nature* **453**, 757-760 (2008).

10  Czaplicki, R. *et al.* Enhancement of second-harmonic generation from metal nanoparticles by passive elements. *Phys. Rev. Lett.* **110**, 093902-093902 (2013).

11  Khurgin, J. B. & Sun, G. Plasmonic enhancement of the third order nonlinear optical phenomena: Figures of merit. *Opt. Exp.* **21**, 27460-27480 (2013).

12  Zheludev, N. I. The road ahead for metamaterials. *Science* **328**, 582-583 (2010).

13  Wegener, M. Metamaterials beyond optics. *Science* **342**, 939-940 (2013).

14  Milton, G. W. *The Theory of Composites*. (Cambridge university press, 2002).

15      Alloatti, L. *et al.* 100 GHz silicon-organic hybrid modulator. *Light Sci. Appl.* **3**, e173-e173 (2014).

16      Bogoni, A., Wu, X. X., Bakhtiari, Z., Nuccio, S. & Willner, A. E. 640 Gbits/s photonic logic gates. *Opt. Lett.* **35**, 3955-3957 (2010).

17      Kwiat, P. G. *et al.* New High-intensity source of polarization-entangled photon pairs. *Phys. Rev. Lett.* **75**, 4337-4341 (1995).

18      Ghahramani, E., Moss, D. J. & Sipe, J. E. Full-band-structure calculation of $2^{nd}$-harmonic generation in odd-period strained $(Si)_n/(Ge)_n$ superlattices. *Phys. Rev. B* **43**, 8990-9002 (1991).

19      Pendry, J. B., Holden, A. J., Robbins, D. J. & Stewart, W. J. Magnetism from conductors and enhanced nonlinear phenomena. *IEEE Trans. Microw. Theory Tech.* **47**, 2075-2084 (1999).

20      George, S. M. Atomic layer deposition: an overview. *Chem. Rev.* **110**, 111-131 (2010).

21      Terhune, R. W., Maker, P. D. & Savage, C. M. Optical harmonic generation in calcite. *Phys. Rev. Lett.* **8**, 404-406; (1962).

22      Shen, Y. R. Surface nonlinear optics: a historical perspective. *IEEE J. Sel. Topics Quantum Electron.* **6**, 1375-1379 (2000).

23      Herman, W. N. & Hayden, L. M. Maker Fringes Revisited - $2^{nd}$-harmonic generation from birefringent or absorbing materials. *J. Opt. Soc. Am. B* **12**, 416-427 (1995).

24      Clark, R. D. *et al.* Vol. 16 291-305 (Electrochemical Society Inc, 2008).

25      Bloember.N. Role of cracks, pores, and absorbing inclusions on laser-induced damage threshold at surfaces of transparent dielectrics. *Appl. Opt.* **12**, 661-664 (1973).

## Acknowledgements

We thank Jacob Khurgin (Baltimore) for discussions, Heike Störmer and Dagmar Gerthsen (KIT) for the TEM images, Johannes Fischer and Andreas Wickberg (KIT) for help regarding the atomic-layer deposition, and Robert Schittny (KIT) for help with figure 3c. We acknowledge support by the DFG-Center for Functional Nanostructures (CFN) through subproject A1.5, by the Karlsruhe School of Optics & Photonics (KSOP), by the Helmholtz International Research School for Teratronics (HIRST), by the European Research Council (ERC Starting Grant 'EnTeraPIC', number 280145), and by the Alfried Krupp von Bohlen und Halbach Foundation.

## Author contributions

L.A. conceived the concept, grew part of the samples, built the experimental setup and created the data in Fig.2. Cl.K. re-measured independently all the samples, performed the polarization measurement and developed the corresponding theory. M.L. supported Cl.K. A.F. developed the ALD recipes. A.F and T.F. grew part of the samples. K..K. and Cl.K. performed the ellipsometry measurements. W.F, J.L. M.W. and Ch.K. supervised the work. The manuscript was written by L.A., Cl.K., J.L., M.W and Ch.K.

# Supplementary Information

# Second-harmonic generation from ABC-nanolaminates grown by atomic layer deposition

**Authors:** Luca Alloatti, Clemens Kieninger, Andreas Froelich, Matthias Lauermann, Tobias Frenzel, Kira Köhnle, Wolfgang Freude, Juerg Leuthold, Martin Wegener and Christian Koos

**Determination of effective bulk metamaterial nonlinear susceptibility tensor elements:** We determined the effective nonlinear susceptibility of the metamaterial by fitting the measured SH power as a function of the pump polarization for a fixed angle of incidence $\vartheta$, Fig.3c. The theoretical framework given by Herman[1] was extended to arbitrary input polarizations, leading to s/p polarized SH power in forward direction:

$$P_2^{s/p} = \frac{2\left((\sin(\varphi))^2 (t_{af}^{(1,p)})^2 + (\cos(\varphi))^2 (t_{af}^{(1,s)})^2\right)^2 (t_{fs}^{(2,s/p)})^2 (t_{sa}^{(2,s/p)})^2}{cA\varepsilon_0 (n_2\cos(\theta_2))^2} d_{eff,s/p}^2 P_1^2 \left(\frac{2\pi}{\lambda}L\right)^2$$

$$\times \frac{\left(\left(\frac{\sin(\Psi)}{\Psi}\right)^2 + (r_{af}^{(2,s/p)})^2 (R^{(s/p)})^2 \left(\frac{\sin(\Phi)}{\Phi}\right)^2 - 2r_{af}^{(2,s/p)} R^{(s/p)} \frac{\sin(\Psi)}{\Psi}\frac{\sin(\Phi)}{\Phi}\cos(2\phi_2)\right)}{1+\left(r_{af}^{(2,s/p)} r_{fs}^{(2,s/p)}\right)^2 + 2r_{af}^{(2,s/p)} r_{fs}^{(2,s/p)}\cos(4\phi_2)}$$

where, following the notation of Herman[1], $t^{(m,s/p)}$ and $r^{(m,s/p)}$ are the standard Fresnel transmission and reflection coefficients, respectively, at the air (*a*) – film (*f*), film (*f*) – substrate (*s*), and substrate (*s*) – air (*a*) interfaces for s/p-polarization at the frequency $m\omega$, $m = 1, 2$. The quantity $c$ denotes the speed of light, $A$ is the spot size of the laser at the focus, $n_m$ is the refractive index of the nonlinear metamaterial at $m\omega$, $P_1$ is the pump power, $L$ is the thickness of the nonlinear metamaterial, $\lambda$ is the vacuum wavelength of the fundamental beam, and $\varphi$ is the pump polarization angle in air with $\varphi = 0°, 90°$ for s and p-polarizations respectively, $\Psi = \phi_1 - \phi_2$, and $\Phi = \phi_1 + \phi_2$ with $\phi_m = \frac{2\pi L}{\lambda} n_m \cos(\vartheta_m)$, where $\vartheta_m$ is given by Snell's law $\sin(\vartheta_m) = \frac{1}{n_m}\sin(\vartheta)$. The $\vartheta$-dependent phase-mismatch $\Psi$ is responsible for the well-known Maker fringes and can be neglected for films as thin as ours. The effective susceptibility for s/p polarized SH generation, $d_{eff}^{s/p}$, takes the following form for $C_{\infty,v}$ space symmetry (here $d_{33} = \frac{1}{2}\chi_{zzz}$, $d_{15} = \frac{1}{2}\chi_{xzx}$, $d_{31} = \frac{1}{2}\chi_{zxx}$):

$$d_{eff}^p = -d_{15}\cos(\vartheta_2)\sin(2\vartheta_1)\sin^2(\varphi') - d_{31}\sin(\vartheta_2)\left(\cos^2(\vartheta_1)\sin^2(\varphi') + \cos^2(\varphi')\right)$$
$$\quad -d_{33}\sin(\vartheta_2)\sin^2(\vartheta_1)\sin^2(\varphi')$$
$$d_{eff}^s = -d_{15}\sin(\vartheta_1)\sin(2\varphi')$$

The quantity $\varphi'$ is the polarization angle of the pump inside the metamaterial,

$$\tan(\varphi') = \frac{t_{a,f}^{(1,p)}}{t_{a,f}^{(1,s)}} \tan(\varphi).$$

The coefficients $R^{s/p}$, which account for the back-reflected SH wave in the nonlinear film, are given by[1]:

$$R^s = \frac{-d_{15}\sin(\vartheta_1)\sin(2\varphi')}{-d_{15}\sin(\vartheta_1)\sin(2\varphi')} = 1$$

$$R^p = \frac{d_{15}\cos(\vartheta_2)\sin(2\vartheta_1)\sin^2(\varphi') - d_{31}\sin(\vartheta_2)\left(\cos^2(\vartheta_1)\sin^2(\varphi') + \cos^2(\varphi')\right)}{-d_{15}\cos(\vartheta_2)\sin(2\vartheta_1)\sin^2(\varphi') - d_{31}\sin(\vartheta_2)\left(\cos^2(\vartheta_1)\sin^2(\varphi') + \cos^2(\varphi')\right)}$$

The absolute SH power was calculated using the photomultiplier responsivity specified by the manufacturer (e.g., $4.5\times10^6$ A/W for the maximum cathode voltage) and was obtained by a least-squares fit, therefore providing the three independent tensor elements.

**Experiments on 14 additional different samples:** We have grown and characterized 14 additional different samples: Herein, the ABCABC… sequence is changed to ACBACB…. While these are mirror images of each other, the atomic-layer-deposition growth of material B on A is not necessarily equivalent to that of B on C. For $N = 12$, $M = 25$, and $M \times (3N) = 900$ growth cycles total, the ACBACB... sample yields twice the SH power of the ABCABC… sample. As the TiO$_2$ layers in Fig. 2 are five times thinner than the other two layers, we have grown samples with $N$ TiO$_2$ cycles, $N'$ Al$_2$O$_3$ cycles, and $N'$ HfO$_2$ cycles, with $N = 12$ and $N' = 4, 6, 7, 8, 9$, while choosing the integer number of periods $M$ such that the total number of growth cycles $M \times (N' + N' + N)$ is closest to 900. Out of these, the sample with $N' = 8$ yields the largest SH power, which is about 35% larger than that for the ACB sample with $N = N' = 12$. We have similarly grown samples in which either the Al$_2$O$_3$ or HfO$_2$ growth cycle number is fixed to $N = 12$ and the other two are varied like $N' = 4, 6, 8, 10$. All of these samples yield smaller SH powers.

**References**

1    Herman, W. N. & Hayden, L. M. Maker fringes revisited - 2$^{nd}$-harmonic generation from birefringent or absorbing materials. *J. of the Optical Society of America B-Opt. Phys.* **12**, 416-427 (1995).